\documentclass[12pt]{article}
\usepackage{amssymb}
\usepackage{amsmath}

\newcommand{\E}{\mathrm{e}}                                
\newcommand{\qed}
           {\mbox{\quad\rule[-1.5pt]{.4em}{1.5ex}}}       

\newcommand{\PF}{\textsc{Proof:\quad}}                    
\newtheorem{claim}{Claim}[section]
\newtheorem{prop}[claim]{Proposition}                     
\newtheorem{thm}[claim]{Theorem}                          
\newtheorem{rem}[claim]{Remark}                           

\newcommand{\beqn}{\begin{eqnarray}}                      
\newcommand{\eeqn}{\end{eqnarray}}                        

\def\OMIT#1{}                                             

\def\th{\theta}
\def\e{\varepsilon}
\def\g{\gamma}
\def\G{\Gamma}
\def\l{\lambda}
\def\p{\partial}
\def\D{\Delta}
\def\bs{\backslash}

\def\a{\alpha}
\def\b{\beta}

\def\Si{\Sigma}

\begin{document}

\title{\textbf{Geometric coupling thresholds \\ in a two-dimensional strip}}
\author{D.~Borisov$^a$, P.~Exner,$^{b,c}$ and R.~Gadyl'shin$^{a,d}$}
\date{}
\maketitle

\begin{quote}
{\small {\em a) Bashkir State Pedagogical University, October
Revolution
\\
\phantom{a) } St.~3a, 450000 Ufa, Russia
\\
b) Nuclear Physics Institute, Academy of Sciences, 25068 \v Re\v
z
\\
\phantom{a) }near Prague, Czechia
\\
c) Doppler Institute, Czech Technical University, B\v rehov{\'a}
7,
\\
\phantom{a) }11519 Prague, Czechia
\\ d) Institute of Mathematics, Ufa Science Center, Russian
\\
\phantom{a) }Academy of Sciences, Chernyshevskogo St., 112,
 450000, \\
\phantom{a) }Ufa, Russia}
\\
\phantom{a) }\texttt{BorisovDI@ic.bashedu.ru},
\texttt{exner@ujf.cas.cz},
\\ \phantom{a) }\texttt{gadylshin@bspu.ru} }
\end{quote}

\begin{quote}
{\small We consider the Laplacian in a strip $\mathbb{R}\times
(0,d)$ with the boundary condition which is Dirichlet except at
the segment of a length $2a$ of one of the boundaries where it
is switched to Neumann. This operator is known to have a
non-empty and simple discrete spectrum for any $a>0$. There is a
sequence $0<a_1<a_2<\cdots$ of critical values at which new
eigenvalues emerge from the continuum when the Neumann window
expands. We find the asymptotic behavior of these eigenvalues
around the thresholds showing that the gap is in the leading
order proportional to $(a\!-\!a_n)^2$ with an explicit
coefficient expressed in terms of the corresponding
threshold-energy resonance eigenfunction.}
\end{quote}


\setcounter{equation}{0}
\section{Introduction}

Spectra of Dirichlet Laplacians in infinitely stretched regions
such as planar strips or layers with local perturbations were
studied recently in numerous papers. The motivation for this
work came from applications in condensed matter physics, and
also from the fact that it was itself an interesting
mathematical problem.

One of the simplest systems of this type is a free quantum
particle confined to a pair of straight parallel strips with
Dirichlet boundary conditions coupled laterally by a window in
the common boundary. If they are of the same width $d$, one can
employ the mirror symmetry and concentrate on the nontrivial
part which is equivalent to the analysis of the Laplacian in a
single strip with the Dirichlet boundary condition switched to
Neumann at a finite segment of one of the boundaries.

Such a system has at least one bound state for any ``window''
length $2a>0$ as it was found in \cite{ESTV} and independently
in \cite{BGRS}. The discrete spectrum is simple, the eigenvalues
$\lambda_n\,,\: n=0,1,\dots$, are continuously decreasing as
functions of $a$ and their number is linear in $a$ up to an
error term as the window is widening \cite{ESTV}. These
properties follow from a simple bracketing argument which allows
to squeeze the eigenvalues between those of a box covering the
``coupled'' part with Dirichlet and Neumann conditions at
$x_1=\pm a$. It shows, in particular, that there are critical
values $a_n\,,\: n=0,1,\dots$, at which new eigenvalues emerge
from the continuum. By \cite{ESTV, BGRS} we have $a_0=0$ while
generally we know only that
 \begin{equation} \label{critpoints}
 a_n \in \left( \frac{nd}{\sqrt{3}}\,,\, \frac{(n+1)d}{\sqrt{3}}
 \right)\,, \quad n=1,2,\dots\,.
 \end{equation}
This tells us nothing about the behavior of the eigenvalues
around the critical points.

The weak coupling asymptotics was studied for the ground state.
A variational method of \cite{EV} yields for small $a$ a
two-sided estimate of $\lambda_0(a)$ between two multiples of
$a^4$. This is indeed the leading term: using the matching
method of \cite{Il, Ga} Popov derived in \cite{Po} the expansion
 \begin{equation}\label{popov}
 \lambda_0(a)=\left(\frac{\pi}{d}\right)^2- \left(
 \frac{\pi^3}{2d^3}
 \right)^2 a^4+\mathcal{O}(a^5)\,.
 \end{equation}
While his argument is not fully rigorous because an estimate of
the error term is missing, the formula itself raises no doubts,
in particular, because of its excellent agreement with the
numerical result of \cite{ESTV}. The result is subtle: recall
that (\ref{popov}) differs substantially from the asymptotics
corresponding to a local change in {\em mixed} boundary
conditions, where the Birman-Schwinger technique is applicable
and the leading term is a multiple of the window width {\em
squared} \cite{EK}.

In the present paper we address the question about the behavior
of the higher eigenvalues $\lambda_n\,,\: n=1,2,\dots$, in the
vicinity of the critical points $a_n$. It is a natural
counterpart of the coupling constant threshold problem of
\cite{KS}. Our main result is that the gap between an eigenvalue
$\lambda_n$ and the continuum is proportional to $(a-a_n)^2$
with a coefficient given explicitly in terms of the
corresponding threshold-energy resonance eigenfunction. This
fits well into the analogy between our problem and spectral
properties of one-dimensional Schr\"odinger operators. The
latter extends to higher dimensions, but the argument becomes
more complicated and we leave that to another paper.

To finish the introduction, let us say something about the
method. We deal with a perturbative problem with respect to the
parameter $\e$ defined as the excess of the window halfwidth $a$
over the critical value $a_n$. Since the latter is positive for
$n\ge 1$, the problem in question is a regular one. This fact
makes it possible to map the problem into an equivalent one with
a small and local perturbation of the equation and the boundary
condition fixed, i.e. independent of $\e$. This is what we are
going to do. We employ the technique introduced in \cite{Ga_TMF}
for calculations of the eigenfunctions for one-dimensional
perturbed Schr\"odinger operator; its advantage is that we
arrive at the sought asymptotic formula in a straightforward and
reasonably simple way.

\setcounter{equation}{0}
\section{The main result}

To formulate the result we need first to introduce some notation
and recall a few simple facts about the problem in question. We
employ Cartesian coordinates, $x=(x_1,x_2)$, in which $\Sigma=
\{\, x\,:\, 0<x_2<d\,\}$. The upper strip boundary is $\Gamma=
\{\, x\,:\, x_2=d\,\}$, while the lower decomposes into the union
of $\gamma^a= \{\, x\,:\, |\,x_1|<a\,,\,x_2=0\,\}$ and $\Gamma^a=
\{\, x\,:\, |\,x_1|> a\,,\,x_2=0\,\}$. The operator $H_a$ we are
going to consider is the Laplacian, $H_a\psi= -\Delta\psi$, in
$L^2(\Sigma)$ with the Dirichlet boundary condition on $\Gamma^a
\cup \Gamma$ and Neumann on $\gamma^a$; it is a well-defined
self-adjoint operator -- cf.~\cite[Chap.~7]{Da}, \cite{DK}.
 \begin{prop}\label{pr2.1}
 The discrete part of $\sigma(H_a)$ is simple and the eigenvalues
 $\lambda_n(a)\,:\: \left(\frac{\pi}{2d}\right)^2< \lambda_0(a)<
 \lambda_1(a)<\cdots< \left(\frac{\pi}{d}\right)^2$ are continuous and
 monotonously decreasing with respect to $a$. There are numbers
 $0=a_0<a_1<a_2 <\cdots$ satisfying (\ref{critpoints}) such that
 \\ [.3em]
 (a) for $a\in(a_{n-1},a_n]$ the operator $H_a$ has exactly $n$
 eigenvalues, \\ [.3em]
 (b) for $a>a_n$ we denote $\e:=a\!-\!a_n$, then the eigenfunction
 $\psi_{n}^{(\e)}$ associated with $\l_n(a)$ has a definite parity
 with respect to $x_1$, namely
 $$ 
\psi_n^{(\e)}(-x_1,x_2)=(-1)^n\psi_n^{(\e)}(x_1,x_2)\,,
 $$ 
(c) for $a=a_n$ the equation $-\Delta\psi= \left(\frac{\pi}{
 d}\right)^2 \psi$ with given boundary conditions has a solution
 $\psi_n^{(0)}\in H^1_\mathrm{loc}(\Sigma)$, unique up to a
 multiplicative constant, which behaves like
 \begin{equation}\label{thresh-asympt}
 \psi_n^{(0)}(x) = c_1 (\pm 1)^n \sin\left(\frac{\pi x_2}{d}\right) +
 \mathcal{O}\left( \E^{-\delta |x_1|} \right)
 \end{equation}
 as $x_1\to \pm\infty$, where $\delta:= {\pi \sqrt{3}/ d}\,$ (in
 what follows we
 set $c_1:=\sqrt{2/\pi}$ for the sake of definiteness).
\end{prop}
\PF Most part was demonstrated in \cite{ESTV}, it remains to
check the claim (c). Any solution can be expanded in terms of
the transverse eigenfunction bases
 $$ 
 \chi_j(x_2):= \sqrt{2\over d}\, \sin\left(\pi j x_2\over d\right)\,,
 \; \phi_j(x_2):= \sqrt{2\over d}\, \sin\left(\pi (2j\!-\!1)
 (d\!-\!x_2)\over d\right)
 $$ 
with $j=1,2,\dots\:$. Since the problem has a mirror symmetry
with respect to $x_1=0$, it is sufficient to discuss the
halfstrip part, $x_1\ge 0$, with the appropriate boundary
condition at $x_1=0$. Let is consider the even case. A solution
of energy $\epsilon \left(\pi\over d\right)^2$ expresses as
 \begin{equation} \label{Ansatz}
 \psi(x) = \sum_{j=1}^\infty c_j\, \E^{q_j(a-x_1)} \chi_j(x_2)\,,
 \quad \psi(x) = \sum_{j=1}^\infty b_j\, {\cosh(p_j x_1)\over
 \cosh(p_j a)}\, \phi_j(x_2)
 \end{equation}
for $|\,x_1|\ge a$ and $|\,x_1|\le a$, respectively, where
$q_j:= {\pi\over d} \sqrt{j^2\!-\epsilon}$ and $p_j:= {\pi\over
d} \sqrt{\left(j-{1\over 2}\right)^2\!-\epsilon}$. The
coefficients in the above relation are determined by the
requirement of smoothness of $\psi$ at the segment $x_1=a$; we
have
 \begin{equation} \label{c-b}
 c_j = \sum_{k=1}^\infty b_k(\chi_j,\phi_k)\,, \quad
 (\chi_j,\phi_k) = {(-1)^{j-k}\over \pi}\, {2j\over j^2\!-\!
 \left(k\!-{1\over 2}\right)^2}\,,
 \end{equation}
and $b=\{b_j\}$ is given as solution of an infinite system of
equations which can be written concisely in the operator form
 \begin{equation} \label{coef-eq}
 Cb=0
 \end{equation}
with
 $$ 
 C_{jk}:= \left\lbrack q_j+p_k\tanh(p_k a) \right\rbrack\,
 (\chi_j,\phi_k)\,.
 $$ 
The odd case is similar, just $\cosh$ and $\tanh$ are replaced
by $\sinh$ and $\coth$, respectively. The allowed values of
$\epsilon$ are those for which a solution to the system
(\ref{coef-eq}) exists.

We know from \cite{ESTV} that the sequence corresponding to an
isolated eigenvalue of $H_a$, i.e. $\epsilon\in \left( {1\over
4}, 1 \right)$, belongs to $\ell^2(j^{-r})$ for any $r\ge 1$,
and that $C=C(a,\epsilon)$ is Hilbert-Schmidt on
$\ell^2(j^{-r})$ with $r$ large enough independently of $a$ and
$\epsilon$. Choosing such an $r$ it is straightforward to check
that $(a,\epsilon) \mapsto C(a,\epsilon)$ is jointly continuous
in the corresponding Hilbert-Schmidt norm. Take $a>a_n$ and
$\epsilon_n(a):= \left( d\over\pi \right)^2 \lambda_n(a)$,
clearly $\epsilon_n(a)\to 1$ as $a\to a_n+$. The said continuity
implies that the equation (\ref{coef-eq}) has for $C(a_n,1)$ a
unique solution in $\ell^2(j^{-r})$, and by (\ref{c-b}) it
determines a sequence $b\in\ell^2(j^{-r})$. Together they yield
the sought threshold resonance solution for $a=a_n$ with the
asymptotics (\ref{thresh-asympt}) following from the first one
of the relations (\ref{Ansatz}). $\;$\qed \vspace{1em}

\begin{rem}\label{rm2.1}
{\rm  The function $\psi^{(0)}_n$ described in
Proposition~\ref{pr2.1} has the following smoothness properties.
It is infinitely differentiable everywhere in $\overline{\Si}$
except the endpoints of the segment $\g^{a_n}$. At these points
the asymptotic formula
\begin{equation}
\psi^{(0)}_n(x) =(\pm 1)^n\a_n\,
r_\pm^{1/2}\sin\frac{\th_\pm}{2}+\mathcal{O}(r_\pm)\,,\quad
r_\pm \to 0\,, \label{6.5}
\end{equation}
is valid, where $(r_\pm,\th_\pm)$ are polar coordinates associated
with the variables $(\pm x_1\!-\!a_n,x_2)$ and $\a_n$ is a some
number (a unique one provided we fix $c_1$). These asymptotics
formulas can be verified in two easy steps. First, one should
extend the function $\psi_n^{(0)}$ to the mirrored strip $\{x:
-d<x_2<0\}$ in the even way. This leads to solution of the
equation $-\D\psi_n^{(0)}= \left(\frac{\pi}{d}
\right)^2\psi_n^{(0)}$ in the double strip $\{x:
|x_2|<d\}\setminus\{x: |x_1|>a, x_2=0\}$ with the Dirichlet
boundary condition at the outer boundary and at the cut $\{x:
|x_1|>a, x_2=\pm0\}$, and it is sufficient to employ the results
established in \cite{NP} for such elliptic problems. }
\end{rem}

Now we can state our main result:
 \begin{thm} \label{main}
 The eigenvalue $\lambda_n(a)$ of $H_a$ with $n\ge 1$ has the
 following asymptotic behavior,
 \begin{equation}\label{ev-asympt}
 \lambda_n(a) = \left(\pi\over d\right)^2 - \mu_n^2 (a\!-\!a_n)^2 +
 \mathcal{O}\left((a\!-\!a_n)^3 \right)
 \end{equation}
 as $a\to a_n+$, where the coefficient is given by
 \begin{equation}\label{M_1}
 \mu_n:=\frac{1}{a_n}
 \int\limits_{\Sigma} \left|\frac{\partial\psi_n^{(0)}}{\partial
 x_1}\right|^2\,dx_1dx_2,
 \end{equation}
 or alternatively by
 \begin{equation}\label{alM_1}
 \mu_n:=\frac{\pi\a_n^2}{4}\,,
 \end{equation}
where $\a_n$ is the number appearing in (\ref{6.5}). The
associated eigenfunction $\psi_n^{(\e)}$ can be normalized in
such a way that it satisfies the relation
 \begin{equation}\label{egf-con}
 \psi^{(\e)}_n=\psi^{(0)}_n+\mathcal{O}(\e)\,,
 \end{equation}
in $H^1\left((-R,R)\times (0,d) \right)$ for any $R>0$ behaving
asymptotically as
 \begin{equation}\label{as-egf}
 \psi^{(\e)}_n(x)=c_1\, (\pm 1)^n \,\mathrm{e}^{-\e \mu_n |x_1|}
 \sin\left(\pi x_2\over d\right)+ \mathcal{O}(\mathrm{e}^{
 -\delta|x_1|})
 \end{equation}
when $x_1\to\pm\infty$, with $\delta$ defined in
Proposition~\ref{pr2.1}.
\end{thm}
\begin{rem}\label{rm2.2}
{\rm The function $\psi_n^{(\e)}$ belongs, of course, to
$L^2(\Sigma)$, but it does not have a limit in this space as
$\e\to0+$ since the norm $\|\psi_n^{(\e)}\|_{L^2(\Sigma)}$
explodes in the limit. Furthermore, $\a_n$ is nonzero for any
$n\ge 1$. This fact  can be easily deduced from the assertion
(\ref{M_1}). Indeed, the assumption $\a_n=0$ implies immediately
that $\psi_n^{(0)}$ is independent on $x_1$. However, this
contradicts to the boundary value problems that $\psi_n^{(0)}$
satisfies to. The coefficient $\a_n$ being nonzero, the formula
(\ref{alM_1}) shows that the asymptotics of each $\l_n$ is
nontrivial in the leading order. }
\end{rem}

\setcounter{equation}{0}
\section{Proof of Theorem 2.2}

The rest of the paper is devoted to the proof of
Theorem~\ref{main}. With the scaling behavior of the spectrum in
mind we can put $d=\pi$ without loss of generality in the
following. Furthermore, the mirror symmetry with respect to
$x_1=0$ makes it possible to consider the halfstrip problem with
the Dirichlet or Neumann condition at the cut.

We need to introduce some notations. Let $\mathbb{R}^2_+=\{x:
x_1>0\}$ be the open right halfplane. As indicated above, we
will work in the halfstrip $\Pi:=\Sigma\cap \mathbb{R}^2_+$,
similarly we introduce $\g_\e:= \g^{a_n+\e}\cap\mathbb{R}^2_+$,
$\G_+:= \G\cap\mathbb{R}^2_+$, and $\G_\e:= \G^{a_n+\e}\cap
\mathbb{R}^2_+$. Moreover, we need a symbol for the cut at the
symmetry axis, $\g:=\{x: x_1=0\,,\, 0<x_2<\pi\}$. Since we
consider a fixed eigenvalue, we shall omit the index $n$ when
there is no danger of misunderstanding.

Let us outline the strategy of the proof. In the first step we
are going to analyze the problem
\begin{align}
(\D+1)u &=m^2 u+f\,,\quad x\in\Pi\,,\label{2.2}
\\
l_x u &=0\,,\quad x\in \g\,, \nonumber
\\
u &=0\,,\quad x\in \G_0\cup\G_+\,,\nonumber
\\
\frac{\p u}{\p x_2}&=0\,,\quad x\in \g_0\,, \nonumber
\end{align}
with a fixed function $f$ at the right-hand side of the
equation. The trace operator $l_x$ in the boundary conditions is
defined as $(l_x u)(x_1,x_2)=u(0,x_2)$ if $n$ is odd and $(l_x
u)(x_1,x_2)=\frac{\p u}{\p x_1}(0,x_2)$ if $n$ is even. The
parameter $m$ is assumed to be complex and to lie in a
(sufficiently small) neighborhood of zero (we indicate this
neighborhood by $\mathcal{D}$). We also suppose that $f$ is an
arbitrary function from $L^2(\Pi)$ with a compact support. We
will construct a solution of the problem (\ref{2.2}),
meromorphic with respect to $m$, with the following asymptotic
behavior far from the cut of the halfstrip,
\begin{equation}\label{3.8}
u(x,m)=c(m)\, \mathrm{e}^{-m x_1}\sin x_2+ \mathcal{O}
(\mathrm{e}^{-\sqrt{3+m^2}x_1})\,, \quad x_1\to+\infty\,,
\end{equation}
where $c(m)$ is a constant determined by the function $f$. In
the second step we will transform the original boundary value
problem for the eigenfunction $\psi^{(\e)}$ to another one with
an equation the coefficients of which depend smoothly on $\e$
and the boundary condition is independent of $\e$. What is
important is that the reformulated problem will be of the form
(\ref{2.2}) with a particular right-hand side $f=f_\e$ for which
we deduce a sufficient explicit representation. Combining the
latter with properties of the solution to (\ref{2.2}) will
finally obtain the announced results concerning $\l_\e$ and
$\psi^{(\e)}$. Since the proof of Theorem~\ref{main} divides in
this way naturally into two steps, we shall discuss them
separately in the two following subsections.

\subsection{Solution of the problem (\ref{2.2})}

As we have indicated our aim is to construct a solution of
(\ref{2.2}) which is meromorphic in $m\in \mathcal{D}$. Let us
say more explicitly what we mean by that. It is easy to see that
there is a unique solution for $m\in\mathcal{D}\cap\{m:
\mathrm{Re}\,m>0\}$ which decays as $x_1\to\infty$. We shall
check that as a function of $m$ it is analytic in
$\mathcal{D}\cap\{m: \mathrm{Re}\,m>0\}$ and extend it to the
remaining part of $\mathcal{D}$. The extension will be for
$\mathrm{Re}\,m<0$ again a solution to (\ref{2.2}) with the
asymptotics (\ref{3.8}), and in addition, it will be meromorphic
in $m\in\mathcal{D}$ with just one simple pole at $m=0$. Of
course, the extension will be bounded at large distances if
$\mathrm{Re}\, m=0$ and will be increasing for $\mathrm{Re}\,
m<0$. Speaking about solutions everywhere in the following we
mean always such analytic continuations.

Recall that a function $F$ with values in some Banach space $X$
is said to be holomorphic in $\mathcal{D}$ if it is
differentiable (in the sense of the norm of $X$) at each point
$m\in\mathcal{D}$. If this function is holomorphic in
$\mathcal{D}$ everywhere except a discrete set of points which
are poles of $F$, i.e. Laurent's series of $F$ at such a point
has at most a finite number of negative terms, then $F$ is said
to be meromorphic.

We introduce some notations. Let $\mathcal{L}(X,Y)$ be the
Banach space of  bounded linear operators from a Banach space
$X$ into a Banach space $Y$. We will use the symbol
$\mathcal{H}(X)$ for the class of functions with values in $X$
which are holomorphic with respect to $m\in \mathcal{D}$, and
$\mathcal{M}(X)$ for the class of meromorphic on
$m\in\mathcal{D}$ functions with values in $X$. For the sake of
brevity we also introduce the notations $\mathcal{H}(X,Y):=
\mathcal{H}(\mathcal{L}(X,Y))$, $\mathcal{M}(X,Y):=
\mathcal{M}(\mathcal{L}(X,Y))$.

We will treat the problem (\ref{2.2}) by the technique
introduced in \cite[Sec.~XVI.4]{SP}. Let $R$ be a fixed number
larger than $a_n$, $\Pi_R:=\Pi\cap\{x: x_1<R\}$, and $g$ a
function from $L^2(\Pi_R)$ which can be also regarded as an
element of $L^2(\Pi)$ if we put it equal to zero for $x_1>R$.
The problem
\begin{equation}\label{3.2}
 (\D+1)v=m^2 v + g\,,\quad x\in\Pi\,,\qquad v=0\,,\quad
 x\in\p\Pi\,,
\end{equation}
can be easily solved by separation of variables, the solution
being
\begin{align}
 \label{3.3} v(x,m)&=
-\sum\limits_{k=1}^\infty\frac{1}{\pi M_k}
 \int\limits_\Pi G_k(x,t,m)g(t)\,d^2t\,,
 \\
 G_k(x,t,m) &=\left(\mathrm{e}^{-M_k|x_1-t_1|}
 -\mathrm{e}^{-M_k(x_1+t_1)}\right)\sin k t_2 \sin k x_2\,,
 \nonumber
\end{align}
where $M_1=m$ and $M_k=\sqrt{k^2\!-1+m^2}$ for $k\ge 2$. For the
sake of brevity we use the notation $d^2t=dt_1\,dt_2$.
Obviously, the formula (\ref{3.3}) is valid for all
$m\in\mathcal{D}$, not only for $\mathrm{Re}\,m>0$. The function
$v$ can be represented as $v=T_1(m)g$, where $T_1(m):
L^2(\Pi_R)\to H^2(\Pi_{\widetilde{R}})$ is a bounded linear
operator for any positive $\widetilde{R}$. It is straightforward
to check that $v\in\mathcal{H}(H^2(\Pi_{\widetilde{R}}))$ and
$T_1(\cdot)\in\mathcal{H}(L^2(\Pi_R),H^2(\Pi_{\widetilde{R}}))$
in $\mathcal{D}$ for any positive $\widetilde{R}$.

In the next step we consider another boundary value problem for
an unknown function $w$, namely
\begin{align}
\D w &= \D v\,, \quad x\in\Pi_R\,, \label{3.6}
\\
w &= 0\,,\quad x\in(\G_+\cup\G_0)\cap\p\Pi_R\,, \nonumber
\\
w &= v\,, \quad x_1=R\,,\nonumber
\\
\frac{\p w}{\p x_2} &= 0\,, \quad x\in\g_0\,, \nonumber
\\
l_x w &=0\,, \quad x\in\g\,. \nonumber
\end{align}
Since $v\in H^2(\Pi_R)$ we have $\D v\in L^2(\Pi_R)$, and thus
the problem (\ref{3.6}) has a unique solution $w\in H^1(\Pi_R)$
-- see, e.g., \cite[Sec.~2.5, Rem.~5.1]{Ld}. Using the standard
theorem on smoothness of solutions of elliptic boundary value
problems we can conclude that $w\in H^2(\Pi_{R,s})$ holds for
each $s>0$, where $\Pi_{R,s}:=\Pi_R\bs D_s$ with $D_s:=\{x:\,
x_2~>~0\,,\newline \;(x_1\!-\!a_n)^2+x_2^2<s^2\}$. Hence we can
define the linear operator $T_2$ by $w=:T_2v$ which is a linear
bounded map from $H^2(\Pi_R)$ into $H^1(\Pi_R)$, and from
$H^2(\Pi_R)$ into $H^2(\Pi_{R,s})$ for each fixed $s$.

Let $\chi_{R}(x_1)$ be an infinitely differentiable mollifier
function equal to one for $x_1\le R-1$ and vanishing for $x_1\ge
R$. We take the two functions considered above and construct $u$
as a smooth interpolation between them,
\begin{equation} \label{interpol}
u(x,m):=\chi_{R}(x_1)w(x,m)+(1\!-\!\chi_{R}(x_1))v(x,m)\,.
\end{equation}
Since $w(x,m)=T_2 T_1(m)g$, we have $w\in
\mathcal{H}(H^1(\Pi_R))\cap \mathcal{H}(H^2(\Pi_{R,s}))$ as a
function of $m$ for each $s>0$. Thus we can introduce the
operator $T_3(m)$ which maps a function $g\in L^2(\Pi_R)$ to the
function $u$ determined by (\ref{interpol}), where $v$ and $w$
are the solutions of (\ref{3.2}) and (\ref{3.6}), respectively,
for this $g$. It easy to see that
$T_3(\cdot)\in\mathcal{H}(L^2(\Pi_R),H^1(\Pi_{\widetilde{R}}))$
and
$T_3(\cdot)\in\mathcal{H}(L^2(\Pi_R),H^2(\Pi_{\widetilde{R},s}))$
in $\mathcal{D}$ for any pair of positive $\widetilde{R}$, $s$.

According to the definition of $v$ and $w$, the function $u$
satisfies all the boundary condition of (\ref{2.2}). Applying
the operator $(\D+1-m^2)$ to this function, we obtain
 \begin{align*}
(\D+1-m^2)u &= g+(w\!-\!v)(\D+1-m^2)\chi_{R}+ \\ &+
2\left(\nabla_x\chi_{R}, \nabla_x(w\!-\!v)
\right)_{\mathbb{R}^2}\,,
 \end{align*}
where we have used in the calculation the equations which $v$
and $w$ satisfy. This result shows that the function $u$ solves
the problem (\ref{2.2}) if and only if $g$ satisfies the
following equation,
\begin{equation}
 g+T_4(m)g=f\,, \label{3.7}
\end{equation}
where
\begin{equation*}
 T_4(m)g:=(w\!-\!v)(\D+1-m^2)\chi_{R}+2\left(\nabla_x\chi_{R},
 \nabla_x(w\!-\!v)\right)_{\mathbb{R}^2}\,;
\end{equation*}
recall that both $w$ and $v$ are obtained from $g$ by actions of
the operators specified above.

\begin{prop}\label{pr3.1}
The operator $T_4(m)$ is compact for any $R>0$ as an element of
$\mathcal{L}(L^2(\Pi_R),L^2(\Pi_R))$ and the function $m\mapsto
T_4(m)$ belongs to $\mathcal{H}(L^2(\Pi_R),L^2(\Pi_R))$ in
$\mathcal{D}$.
\end{prop}
\PF We denote
\begin{align*}
T_{41}(m)g:=&(w\!-\!v)(\D+1\!-\!m^2)\chi_{R},
\\
T_{42}(m)g:=&2\left(\nabla_x\chi_{R},
\nabla_x(w\!-\!v)\right)_{\mathbb{R}^2} +(w\!-\!v)\D \chi_{R}\,.
\end{align*}
Using the described properties of $T_1(m)$ and $T_2$ it is easy
to see that $T_{41}(m)$ is a bounded linear map from
$L^2(\Pi_R)$ into $H^1(\Pi_R)$. The operator-valued function
$T_{41}(\cdot)$ belongs to $\mathcal{H}(L^2(\Pi_R),
H^1(\Pi_R))$. The function $\chi'_R$ is smooth and its support
lies in $\Pi_R\backslash\overline{\Pi}_{R-1}$. Using this
together with the fact that the function $w\!-\!v$ belongs to
$H^2(\Pi_R\backslash\overline{\Pi}_{R-1})$, we conclude that
$T_{42}(m): L^2(\Pi_R)\to H^1(\Pi_R)$ is a linear bounded
operator belonging to $\mathcal{H}(L^2(\Pi_R), H^1(\Pi_R))$ as a
function of $m$. Hence $T_4(\cdot)=T_{41}(\cdot)+T_{42}(\cdot)
\in\mathcal{H}(L^2(\Pi_R),H^1(\Pi_R) )$, and therefore $T_4(m)$
is compact when considered as an operator from $L^2(\Pi_R)$ to
$L^2(\Pi_R)$, belonging to $\mathcal{H}(L^2(\Pi_R),L^2(\Pi_R))$
w.r.t. $m$. $\;$\qed \vspace{1em}

Proposition~\ref{pr3.1} shows that the equation (\ref{3.7}) can
be studied using Fredholm theorems. This will help us to solve
our original problem: to construct the solution of (\ref{2.2})
we have to solve the equation (\ref{3.7}), then by the procedure
described above its solution gives rise to the solution $u$ of
the boundary value problem (\ref{2.2}): $u=T_3(m)g$.

\begin{prop}\label{th3.1}
The problems (\ref{3.7}) and (\ref{2.2}) are equivalent: to each
solution $g$ of (\ref{3.7}) there is a unique solution
$u=T_3(m)g$ of (\ref{2.2}), and vice versa, for each solution of
(\ref{2.2}) there exists a unique $g$ solving (\ref{3.7}) such
that $u=T_3(m)g$. The equivalence holds for any $m\in
\mathcal{D}$.
\end{prop}
\PF The first part has been proved above, it remains to check
invertibility of the operator $T_3$. Let $u$ be a solution of
the problem (\ref{2.2}). Notice that the function $u$ is
infinitely differentiable outside the support of $f$ since there
it is a solution of a homogeneous equation. We have to construct
the solution $g$ of (\ref{3.7}) such that $u=T_3 g$. Let us
first determine the functions $v$ and $w$. By $U$ we denote a
solution of the following problem,
\begin{align}\label{3.9}
\D U &=0\,, \quad x\in\Pi_R\,,
\\
U &= 0\,, \quad x\in \p\Pi_R\bs\overline{\g_0\cup\g}\,,
\nonumber
\\
 U &= u\,,\quad x\in\g_0\cup\g\,, \nonumber
\end{align}
which is unique in $H^1(\Pi_R)$, and moreover, it belongs to
$C^\infty(\Pi_R)$ by a standard result on the smoothness of
solutions of elliptic equations. We set
\begin{align*}
v(x,m) &:= u(x,m)-\chi_{R}(x_1)U(x,m)\,,
\\
w(x,m) &:= u(x,m)+(1\!-\!\chi_{R}(x_1))U(x,m)\,.
\end{align*}
One checks easily that the functions $u=\chi_{R}w
+(1\!-\!\chi_{R})v$, $w$ and $v$ satisfy all the required
boundary condition and that $\D w=\D v$ holds in $\Pi_R$. Now it
suffices to use the equation of the problem (\ref{3.2}) to
determine the function $g$ by
 \begin{align*}
 g(x,m) &:= (\D+1-m^2)v(x,m) \\
 &= (\D+1-m^2)(u(x,m)-\chi_{R}(x_1) U(x,m)) \\
 &= f(x)-(\D+1-m^2)(\chi_{R}(x_1)U(x,m))\,.
 \end{align*}
Let us check that this $g$ solves (\ref{3.7}). Using the
definition of $U$, $v$, and $w$, we compute directly the action
of $T_4(m)$ on $g$ obtaining
 \begin{align*}
 T_4(m)g &= U(\D+1-m^2)\chi_{R} +2(\nabla_x \chi_{R},\nabla_x U) \\
 &= U(\D+1-m^2)\chi_{R}+2(\nabla_x \chi_{R},\nabla_x U)+\chi_{R}\D U \\
 &= (\D+1-m^2)(\chi_{R}U)\,.
 \end{align*}
The last two relations show that $g$ is a solution of the
equation (\ref{3.7}). Let us check the uniqueness: suppose that
there are two solutions $g_1$ and $g_2$ of (\ref{3.7}) leading
to the same function $u^{(1)}=u^{(2)}$. Then $g:=g_1\!-\!g_2\ne
0$ gives rise to $u:=u^{(1)}\!-\!u^{(2)}=0$. Let $v$ and $w$ be
the solutions of (\ref{3.2}) and (\ref{3.6}), respectively,
associated with $u$, and put $U:=w\!-\!v$. Then it easy to see
that this $U$ solves (\ref{3.9}), where the boundary condition
on $\g_0\cup\g$ is homogeneous. Such a solution of (\ref{3.9})
is unique, $U=0$. Thus $w=v=u=0$ holds in $\Pi_R$, and therefore
$g=(\D+1-m^2)v=0$ in $\Pi_R$, which is a contradiction. $\;$\qed
\vspace{1em}

The solution of eq.~(\ref{3.7}) depends on $m$, i.e. $g=g(x,m)$.
Our next aim is to clarify a nature of this dependence and to
look what this implies for the solution of the problem
(\ref{2.2}). We employ the following result borrowed from
\cite[Sec.~XV.7]{SP}:
\begin{thm}\label{thsp}
Let $\mathcal{D}$ be an open connected domain of the complex
plane of the variable $m$ and $\{T(m):\, m\in\mathcal{D}\}$ be a
bounded holomorphic family of compact operators from the Banach
space $X$ into itself. Moreover, assume that there exists a
point $m_*\in D$ such that $(I+T(m_*))^{-1}\in
\mathcal{L}(X,X)$. Then $m\mapsto (I+T(m))^{-1}$ is a
meromorphic function in $\mathcal{D}$ with values in
$\mathcal{L}(X,X)$.
\end{thm}

In our case the Banach space mentioned in Theorem~\ref{thsp} is
$L^2(\Pi_R)$ and $T(m)=T_4(m)$. The existence of $m_*$ is easy
to establish: since the equation (\ref{3.7}) is equivalent to
the boundary-value problem (\ref{2.2}) by
Proposition~\ref{th3.1}, it is sufficient to prove the existence
of $m_*$ for which the problem (\ref{2.2}) has no nontrivial
solution for $f=0$ and $m=m_*$ with the asymptotics (\ref{3.8}).
This is true for $m_*>0$ which is sufficiently small, because
assuming the contrary would lead us to the conclusion that
$H_{a_n}$ has the eigenvalue $(1-m_*^2)$, and this in turn would
contradict to the claims (a) and (c) of Proposition~\ref{pr2.1}.
The compactness of $T_4(m)$ and its holomorphic dependence on
$m$ follow from Proposition~\ref{pr3.1}.

Thus we may apply Theorem~\ref{thsp} to  the equation
(\ref{3.7}). We implies that $(I+T_4(m))^{-1}$ exists and it is
meromorphic as an operator-valued function in $\mathcal{D}$, $\;
(I+T_4(\cdot))^{-1}\in \mathcal{M}(L^2(\Pi_R),L^2(\Pi_R))$.
Poles of this function are the values of $m$ for which the
equation (\ref{3.7}) with the vanishing right-hand side has a
nontrivial solution. The value $m_0:=0$ has this property as it
follows from the claim (c) of Proposition~\ref{pr2.1}. Let
$\phi_0$ be a solution of the equation (\ref{3.7}) for $f=0$ and
$m=m_0$. The function $\phi_0$ is unique up to a multiplicative
constant, the uniqueness being implied by
Proposition~\ref{pr2.1}(c) and Proposition~\ref{pr3.1}. The
remaining ambiguity is removed if we set
$\psi^{(0)}=T_3(0)\phi_0$. Making the domain $\mathcal{D}$
smaller if necessary we can achieve that zero is the only pole
of the function $(I+T_4(\cdot))^{-1}$ contained in
$\mathcal{D}$. In such a case the solution $g=(I+T_4(m))^{-1}$
of the equation (\ref{3.7}) can be for any nonzero
$m\in\mathcal{D}$ represented as
\begin{equation}\label{3.15}
g=g(m)=\frac{g_{-k}}{m^k}+\widetilde g(m)\,,
\end{equation}
where the integer $k\ge 1$ is the order of the pole, the
functions $g_{-k},\,\widetilde g(m)$ belong to $L^2(\Pi_R)$ for
nonzero $m\in\mathcal{D}$, and $\widetilde g(\cdot)\in
\mathcal{M}(L^2(\Pi_R))$ may have a pole at zero of order not
exceeding $k\!-\!1$.

Let us stress that Theorem~\ref{thsp} says nothing about the
order of this pole. Next we are going to prove that the pole in
(\ref{3.15}) is simple, i.e. $k=1$. Substituting (\ref{3.15})
into (\ref{3.7}) and comparing the leading terms in $m$ we see
that $g_{-k}=K_0[f]\phi_0$, where $K_0[f]$ is a constant
depending on $f$. The representation (\ref{3.15}) in turn yields
the following expression for $u$:
\begin{equation}\label{3.16}
u=T_3(m)g=\frac{K_0[f]\psi^{(0)}}{m^k}+\widetilde u(m)\,.
\end{equation}
Here $\widetilde u(\cdot)$ is a function belonging to
$\mathcal{M}(H^1(\Pi_{\widetilde{R}}))\cap\mathcal{M}(H^2(\Pi_{\widetilde{R},s}))$
for any positive $\widetilde{R}$, $s$ which again may have pole
at zero of order not exceeding $k\!-\!1$. Multiplying the
equation in the problem (\ref{2.2}) by $\psi^{(0)}$ and
integrating by parts we see that
\begin{equation}\label{3.17}
\begin{aligned}
\int\limits_{\Pi_{\widetilde{R}}} f\psi^{(0)}\,d^2x=
-m^2\int\limits_{\Pi_{\widetilde{R}}}\psi^{(0)}
u\,d^2x+\int\limits_0^\pi\left(\psi^{(0)}\frac{\p u}{\p x_1}-
u\frac{\p \psi^{(0)}}{\p
x_1}\right)\bigg|_{x_1=\widetilde{R}}\,dx_2.
\end{aligned}
\end{equation}
For $d$ sufficiently large we have
\begin{align*}
&u\big|_{x_1=\widetilde{R}}=\sum\limits_{j=1}^{\infty}C_j(m)\sin
j x_2\,,\quad
\psi^{(0)}\big|_{x_1=\widetilde{R}}=\sum\limits_{j=1}^{\infty}c_j\sin
j x_2\,,
\\
&\frac{\p u}{\p
x_1}\bigg|_{x_1=\widetilde{R}}=-\sum\limits_{j=1}^{\infty}
\sqrt{j^2\!-\!1\!+\!m^2}\, C_j(m)\sin j x_2\,,
\\
&\frac{\p\psi^{(0)}}{\p
x_1}\bigg|_{x_1=\widetilde{R}}=-\sum\limits_{j=2}^{\infty}c_j\sqrt{j^2\!-\!1}
\,\sin j x_2\,,
\end{align*}
where the functions $C_j$ in view of (\ref{3.16}) behave as
\begin{equation*}
C_j(m)=\frac{K_0[f]c_j}{m^{k}}+\mathcal{O}(m^{-k+1})\,.
\end{equation*}
Combining the above relations and using the normalization
condition $c_1=\sqrt{2/\pi}$ we deduce that
\begin{align*}
\int\limits_0^\pi\left(\psi^{(0)}\frac{\p u}{\p x_1}- u\frac{\p
\psi^{(0)}}{\p x_1}\right)\bigg|_{x_1=\widetilde{R}}\,dx_2 =
-\frac{K_0[f]}{m^{k-1}}+ \mathcal{O}(m^{-k+2})\,.
\end{align*}
The first integral at the right hand of (\ref{3.17}) behaves as
\begin{equation*}
\begin{aligned}
-m^2\int\limits_{\Pi_{\widetilde{R}}}\psi^{(0)} u\,d^2x=
-\frac{K_0[f]}{m^{k-2}}\int\limits_{\Pi_{\widetilde{R}}}|\psi^{(0)}|^2\,d^2x+
\mathcal{O}(m^{-k+3}).
\end{aligned}
\end{equation*}
in the limit $m\to 0$. Substituting the expressions obtained
above into (\ref{3.17}) and comparing the coefficients at the
powers of $m$, we conclude that $k=1$, and furthermore, that
\begin{equation}
K_0[f]=-\int\limits_{\Pi_{\widetilde{R}}}f\psi^{(0)}\,d^2x
=-\int\limits_{\Pi}f\psi^{(0)}\,d^2x,\label{3.19}
\end{equation}
To get the last relation we have used the fact that $f$ has a
compact support by assumption which therefore lies in
$\Pi_{\widetilde{R}}$ for $\widetilde{R}$ sufficient large.

Next we denote $A(m)=T_3(m)(I+T_4(m))^{-1}$. It follows from the
relations (\ref{3.16}), (\ref{3.19}) that
\begin{equation}\label{3.21}
\begin{aligned}
{}&u=A(m)f=: A_0(m)f+A_1(m)f=K_0[f]\frac{\psi^{(0)}}m+A_1(m)f,
\\
{}& A_1(\cdot)\in
\mathcal{H}(L^2(\Pi_R),\mathcal{H}(H^1(\Pi_{\widetilde{R}})) \cap
\mathcal{H}(L^2(\Pi_R), H^2(\Pi_{\widetilde{R},s})),
\end{aligned}
\end{equation}
where $\widetilde{R}$ and $s$ are arbitrary positive.

It is convenient to summarize all the conclusions made above in
a single theorem which represents the main result of this
subsection.
\begin{thm}\label{th3.3}
Let $f\in L^2(\Pi_R)$, and assume that $\widetilde{R}$, $s$ are
arbitrary positive numbers, $m\in\mathcal{D}$. Then for $m\in
\mathcal{D}\setminus\{0\}$ there exists a unique solution of the
boundary value problem (\ref{2.2}) given by $u=A(m)f$. As a
function on $m$ this solution belongs to
$\mathcal{M}(H^1(\Pi_{\widetilde{R}}))\cap
\mathcal{M}(H^2(\Pi_{\widetilde{R},s}))$. The neighborhood
$\mathcal{D}$ can be chosen in such a way that the function
$A(\cdot)\in
\mathcal{M}(L^2(\Pi_R),H^1(\Pi_{\widetilde{R}}))\cap
\mathcal{M}(L^2(\Pi_R),H^2(\Pi_{\widetilde{R},s}))$ has just one
pole $m_0=0$ which is simple. The operator $A(m)$ can be
decomposed into the sum $A(m)=A_0(m)+A_1(m)$, where the
operators $A_i(m)$ are defined by (\ref{3.21}). Finally, the
relation (\ref{3.19}) holds true.
\end{thm}


\subsection{The asymptotic analysis}

To find the behavior of the quantity $\l_n(a)$ around the
threshold we will analyze $m_\e$ defined as
$m_\e^2:=1\!-\!\l_n(a_n+\e)$, which satisfies $m_\e\to0$ as
$\e\to0$. Let $y=(y_1,y_2)$ be the Cartesian coordinates of a
point in the strip. The boundary value problem problem for the
eigenfunction $\psi^{(\e)}$ can be then written as
\begin{align}
(\D_y+1)\psi^{(\e)}&=m_\e^2\psi^{(\e)}\,,\quad
y\in\Pi,\label{6.3}
\\
l_y \psi^{(\e)} &=0\,,\quad y\in \g\,, \nonumber
\\
\psi^{(\e)} &=0\,,\quad y\in \G_\e\cup\G_+\,,\nonumber
\\
\frac{\p \psi^{(\e)}}{\p y_2}&=0\,,\quad y\in \g_\e\,. \nonumber
\end{align}
As we have announced, we want to get rid of the $\e$-dependent
boundary condition passing to one independent on $\e$, while the
corresponding equation will have coefficients which depend
smoothly on $\e$. To construct the appropriate transformation we
use an infinitely differentiable mollifier function $\chi$ of
the variable $y_1$ which is equal to one for $y_1\in[\b_2,\b_3]$
and vanishes for $y_1\le \b_1$ and $y_1\ge \b_4$. Here $\b_i$,
$i=1,\dots,4$, are positive constants, $\b_1<\b_2<\b_3<\b_4$,
such that $\b_2<a_n\!+\!\e<\b_3$ holds for all $\e$ from a fixed
neighborhood of zero. We consider the function
$x_1:\,\mathbb{R}_+\to \mathbb{R}_+$ defined as
\begin{equation*}
 x_1:\; x_1(y_1,\e)=y_1-\e\,\chi(y_1)\,.
\end{equation*}
It is easy to see, in particular, that $x_1=a_n$ as
$y_1=a_n+\e$. Taking the first two derivatives we get
\begin{equation*}
\frac{d x_1}{dy_1}=1-\e\,\chi'(y_1)\,,\quad \frac{d^2
x_1}{dy_1^2}=-\e\,\chi''(y_1)\,,
\end{equation*}
from where it follows that for all sufficiently small $\e$ the
first derivative of $x_1$ w.r.t. $y_1$ is nonzero everywhere in
$\mathbb{R}_+$. Thus the map $x_1$ is bijective and can be used
to define a change of the variable, $y_1\mapsto x_1(y_1,\e)$.
The problem (\ref{6.3}) expressed in the variables
$x=(x_1,x_2)$, $x_2:=y_2$, becomes
\begin{align}
(\Delta_x+\e L_\e+1)\psi^{(\e)} &= m_\e^2 \psi^{(\e)} \,,\quad
x\in\Pi\,, \label{2.3}
\\
l_x \psi^{(\e)} &= 0\,,\quad x\in \g\,, \nonumber
\\
\psi^{(\e)} &= 0\,,\quad x\in \G_0\cup\G_+\,,\nonumber
\\
\frac{\p \psi^{(\e)}}{\p x_2} &= 0\,,\quad x\in \g_0\,.
\nonumber
\end{align}
The operator $L_\e$ appearing in the transformed equation is
defined by
\begin{align}
& L_\e:=b_{11}(x_1,\e)\frac{\p^2}{\p x_1^2}+ b_1(x_1,
\e)\frac{\p}{\p x_1}\,,\label{2.4}
\\
&b_{11}(x_1,\e):=-2\chi'(y_1(x_1,\e))+\e(\chi'(y_1(x_1,\e)))^2\,,
\nonumber
\\
& b_1(x_1,\e):=-\chi''(y_1(x_1,\e))\,.\nonumber
\end{align}
The functions $b_{11}$ and $b_{1}$ are obviously infinitely
differentiable, they vanish for
$y_1\not\in[\b_1,\b_2]\cup[\b_3,\b_4]$ and satisfy in the limit
$\e\to0$ the asymptotic formulae
\begin{equation*}
b_{11}(x_1,\e)=-2\chi'(x_1)+\mathcal{O}(\e),\quad
b_{1}(x_1,\e)=-\chi''(x_1)+\mathcal{O}(\e)
\end{equation*}
uniformly in the variable $x_1$.

Now we can proceed to the calculation of $m_\e$ and
$\psi^{(\e)}$. Proposition~\ref{pr2.1} ensures that the
eigenfunction and eigenvalue exist. The function $\psi^{(\e)}$
decays as $|x_1|\to\infty$ and $\l_n(a_n+\e)$ is real. These two
facts imply that $m_\e$ is real and positive. It is obvious that
$\psi^{(\e)}$ solves the problem (\ref{2.2}) for $m=m_\e$ and
$f=-\e L_\e\psi^{(\e)}$. This allows us to seek for the solution
of (\ref{2.3}) in the form
\begin{equation}
\psi^{(\e)}=A(m_\e)f_\e\,,\label{6.7}
\end{equation}
where the function $f_\e$ is assumed to an unknown element of
$L^2(\Pi_R)$ with $R>\b_4$. This may appear strange at a glance,
because we know that $f_\e=-\e L_\e\psi^{(\e)}$, however, we
want to obtain another formula for $f_\e$ not involving
$\psi^{(\e)}$. The function $\psi^{(\e)}$ defined by (\ref{6.7})
satisfies all the boundary condition of the problem (\ref{2.3}).
In order to be a solution of (\ref{2.3}), it is necessary and
sufficient for $\psi^{(\e)}$ to be a solution of the
corresponding equation. Substituting into the  latter the
formula for $\psi^{(\e)}$, we arrive at the equation for $f_\e$
which reads
\begin{equation}\label{2.12}
(I+\e  L_\e A(m_\e))f_\e=0\,.
\end{equation}
In view of (\ref{3.21}) we have
\begin{equation}\label{6.4}
A(m_\e)f_\e=K_0[f_\e]\frac{\psi^{(0)}}{m_\e}+A_1(m_\e) f_\e\,.
\end{equation}
We substitute this representation for $A(m_\e)f_\e$ into
(\ref{2.12}) obtaining
\begin{equation}\label{2.13}
f_\e+\frac{\e}{m_\e}K_0[f_\e] L_\e\psi^{(0)}+ \e L_\e
A_1(m_\e)f_\e=0\,.
\end{equation}
It is clear that $L_\e A_1(\cdot)\in \mathcal{H}(L^2(\Pi_R),
L^2(\Pi_R))$ and as a function of $(m,\e)$ the operator $L_\e
A_1(m)$ is jointly continuous. Thus for sufficiently small $\e$
the inverse operator $B(m,\e)=(I+\e L_\e A_1(m))^{-1}:
L^2(\Pi_R)\to L^2(\Pi_R)$ exists and converges to the identity
map as $\e\to0$ uniformly in $m$. It is also obvious that
$B(\cdot,\e)\in\mathcal{H}(L^2(\Pi_R),L^2(\Pi_R))$. Applying now
$B(m_\e,\e) $ to the equation (\ref{2.13}) we find
\begin{equation}\label{2.17}
f_\e+\frac{\e}{m_\e}K_0[f_\e]B(m_\e,\e) L_\e\psi^{(0)}=0\,.
\end{equation}
Applying further $K_0$ to (\ref{2.17}), we get one more
equation,
\begin{equation*}
K_0[f_\e]+\frac{\e}{m_\e}K_0[f_\e]K_0[B(m_\e,\e)
 L_\e\psi^{(0)}]=0\,.
\end{equation*}
Notice that $K_0[f_\e]$ can not be zero, because otherwise
(\ref{2.17}) would imply $f_\e=0$ which yields $\psi^{(\e)}=0$.
The last equation induces the relation
\begin{equation}\label{2.18}
m_\e=-\e K_0[B(m_\e,\e) L_\e\psi^{(0)}]\,,
\end{equation}
which can be regarded as an equation for $m_\e$. It is easy to
see that $B(\cdot,\e) L_\e\psi^{(0)}\in\mathcal{H}(L^2(\Pi_R))$
and the function $(m,\e) \mapsto B(m,\e) L_\e\psi^{(0)}$ is
jointly continuous. This immediately implies that $K_0[B(m,\e)
L_\e\psi^{(0)}]$ is a holomorphic with respect to $m$ and
jointly continuous as a function of $(m,\e)$. Consequently, for
all $\e$ small enough the estimate
\begin{equation*}
\e\left|[K_0[B(m,\e) L_\e\psi^{(0)}]\right|<|\,m|\,,\quad
m\in\p\mathcal{D}\,,
\end{equation*}
holds true. Using this inequality in combination with the
Rouch\'{e} theorem we conclude that the functions
$h_1:\,h_1(m)=m$ and $h_2:\,h_2(m)=m+[K_0[B(m,\e)
L_\e\psi^{(0)}]$ have the same numbers of zeroes inside
$\mathcal{D}$. This means that the equation (\ref{2.18}) has a
unique solution in $\mathcal{D}$ for all sufficient small $\e$.
On the other hand, due to Proposition~\ref{pr2.1} we know that
there is a root of equation (\ref{2.18}) that converges to zero,
namely
$$ m_\e=\sqrt{\l_n(a_n\!+\!\e)-1}>0\,. $$
Consequently, it is the only root of (\ref{2.18}). Thus the
function
\begin{equation}\label{6.1}
f_\e=-\e B(m_\e,\e) L_\e\psi^{(0)}\,,
\end{equation}
where $m_\e$ is the solution of (\ref{2.18}), solves the
equation (\ref{2.12}). It means that the function $f_\e$ defined
by (\ref{6.1}) gives rise to the eigenfunction
$\psi^{(\e)}=A(m_\e)f_\e$ corresponding to the eigenvalue
$\l_n(a_n\!+\!\e)=1-m_\e^2$. In fact, we have also proved that
there is just one value of $m=m_\e$ tending to zero as $\e\to0$,
for which the boundary value problem (\ref{6.3}) has a
nontrivial solution. This solution decays as $|x_1|\to\infty$,
i.e. there are no non-decaying or even increasing solutions.

The equation (\ref{2.18}) allows us to calculate the asymptotic
expansion for $m_\e$. Since $B(m,\e) L_\e\psi^{(0)}$ is
holomorphic with respect to $m$ and jointly continuous in
$(m,\e)$, and since $m_\e$ tends to zero as $\e\to0$, then by
the equation (\ref{2.18}) we have
\begin{align*}
m_\e&=\e\mu+O(\e m_\e)\,, \quad \e\to0\,,
\\
\mu&=-K_0[ L_0\psi^{(0)}]\,,
\\
 L_0&=-2\chi'(x_1)\frac{\p^2}{\p x_1^2}
-\chi''(x_1)\frac{\p}{\p x_1}\:.
\end{align*}
Combining these relations, we can rewrite the error term as
follows,
\begin{equation}\label{2.19}
m_\e=\e\mu+O(\e^2)\,, \quad \e\to0\,.
\end{equation}
Next we want to express the coefficient $\mu$:
\begin{equation}\label{4.2}
\mu=-\int\limits_{\Pi}\psi^{(0)}\left(2 \chi'(x_1)\frac{\p^2
}{\p x_1^2}+\chi''(x_1)\frac{\p}{\p
x_1}\right)\psi^{(0)}\,d^2x\,.
\end{equation}
Using the equation which the function $\psi^{(0)}$ satisfies we
find
\begin{equation*}
\left(2\chi'(x_1)\frac{\p^2 }{\p x_1^2}+\chi''(y_1)\frac{\p}{\p
x_1}\right)\psi^{(0)}=(\D+1)\left(\chi(x_1)\frac{\p\psi^{(0)}}{\p
x_1}\right)\,.
\end{equation*}
Integrating then by parts and taking into account properties of
$\psi^{(0)}$ together with the definition of the mollifier
$\chi$ we have
\begin{align*}
\mu&=-\lim\limits_{s\to0}\int\limits_{\Pi\bs
D_s}\psi^{(0)}(\D+1)\left(\chi(x_1)\frac{\p\psi^{(0)}}{\p
x_1}\right)\,d^2x=
\\
&=\lim\limits_{s\to0}\int\limits_{\p D_s} \left(
\psi^{(0)}\frac{\p^2\psi^{(0)}}{\p r\p
x_1}-\frac{\p\psi^{(0)}}{\p x_1} \frac{\p\psi^{(0)} }{\p
r}\right)\,ds\,.
\end{align*}
In order to evaluate the last integral along $\p D_s$ we replace
$\psi^{(0)}$ by its asymptotics (\ref{6.5}) and pass to limit
$s\to0$ obtaining
\begin{equation*}
\mu=\frac{1}{2}\a^2\int\limits_0^\pi\sin^2\frac{\th_+}{2}\,d\th_+
=\frac{\pi\a^2}{4}\,,
\end{equation*}
which proves (\ref{alM_1}) with $\mu=\mu_n$. In the same way we
get
\begin{align*}
0&=\int\limits_{\Pi}x_1\frac{\p\psi^{(0)}_n}{\p
x_1}(\D+1)\psi^{(0)}_n\,d^2x=a_n\frac{\pi\a^2}{4}+
2\int\limits_{\Pi}\psi^{(0)}_n\frac{\p^2\psi^{(0)}_n }{\p
x_1^2}\,d^2x=
\\
&=a_n \mu_n-2\int\limits_{\Pi}\left|\frac{\p\psi^{(0)}_n}{\p x_1
}\right|^2\,d^2x\,,
\end{align*}
which yields the representation (\ref{M_1}) for $\mu$,
\begin{equation*}
\mu_n=\frac{2}{a_n}\int\limits_{\Pi}\left|\frac{\p\psi^{(0)}_n}{\p
x_1 }\right|^2\,d^2x=\frac{1}{a_n}\int\limits_{\Si}\left|
\frac{\p\psi^{(0)}_n}{\p x_1}\right|^2\,d^2x\,.
\end{equation*}
In the last relation we employed the fact that $\psi^{(0)}_n$
has a definite parity.

Let us finally pass to discussion of the eigenfunction. In view
of (\ref{6.4}), (\ref{6.1}) we find that $\psi^{(\e)}$ is equal
to
\begin{equation*}
A(m_\e)f_\e=-\frac{\e K_0[B(m_\e,\e)
L_\e\psi^{(0)}]}{m_\e}\psi^{(0)}- \e A_1(m_\e) B(m_\e,\e)
L_\e\psi^{(0)}\,.
\end{equation*}
Using now the equation (\ref{2.18}) together with the fact that
the function $A_1(m) B(m,\e) L_\e\psi^{(0)}$ is holomorphic and
continuous as a function of the respective variables, we
conclude that the relation
\begin{equation*}
\psi^{(\e)}=\psi^{(0)}-\e A_1(m_\e) B(m_\e,\e) L_\e\psi^{(0)}
\end{equation*}
is valid in $H^1(\Pi_R)$ for each $R$. This yields the relation
(\ref{egf-con}) and the asymptotic behavior (\ref{as-egf})
concluding thus the proof of Theorem~\ref{main}.

\subsection*{Acknowledgment}

D.B. is grateful for the hospitality extended to him at NPI AS
where a part of this work was done. The research has been
partially supported by GA AS under the contract \#1048101. The
first and the third authors were partially supported by Russian
Fund of Basic Research -- Grants 00-15-96038, 02-01-00693 (D.B.)
and 02-01-00768 (R.G.) as well as by the Ministry of Education
of the Russian Federation -- Grant E00-1.0-53.


\providecommand{\bysame}{\leavevmode\hbox
to3em{\hrulefill}\thinspace}

\end{document}